\documentclass[twocolumn,secnumarabic,amssymb, nobibnotes, aps, prd]{revtex4-1}

\usepackage[utf8]{inputenc}
\usepackage{amsmath}
\usepackage{amsfonts}
\usepackage{amssymb}
\usepackage{multirow}
\usepackage{lipsum}
\usepackage{epsfig}
\usepackage{here}
\usepackage{fdsymbol}
\usepackage{color}
\usepackage{ulem}
\usepackage{graphicx}
\usepackage[caption=false]{subfig}
\begin{document}
%Flujo en junturas X de sección circular y cuadradas

\title{%The double positive effect of the swimming strategy of {\it E. coli} bacteria in a flow
Effect of motility on the transport of bacteria populations through a porous medium}

\author{Adama Creppy}%
\email{adama.creppy@u-psud.fr}
\affiliation{Univ Paris-Sud, CNRS, F-91405. Lab FAST, B\^at 502, Campus Univ, Orsay, F-91405 (France).}

\author{Eric Cl{\'e}ment}%
\email{eric.clement@upmc.fr}
\affiliation{Physique et M{\'e}canique des Milieux H{\'e}t{\'e}rog{e}nes (UMR 7636 ESPCI /CNRS /Univ.~P.M.~Curie /Univ.~Paris-Diderot), 10 rue Vauquelin, 75005 Paris, France.}

\author{Carine Douarche}
\email{carine.douarche@u-psud.fr}
\affiliation{Laboratoire de Physique des Solides, CNRS, Universit{\'e} Paris Sud, Universit{\'e}  Paris--Saclay, 91405 Orsay cedex, France}

\author{Maria Veronica D'Angelo}
\email{veronica.dangelo@gmail.com}
\affiliation{Universidad de Buenos-Aires, Facultad de Ingenier\'{\i}a, GMP-LIA-FMF, CONICET, Paseo Col\'{o}n 850, 1063, Buenos Aires (Argentina).}

\author{Harold Auradou}
\email{auradou@fast.u-psud.fr}
\affiliation{Laboratoire FAST, Univ. Paris Sud, CNRS, Universit{\'e} Paris-Saclay, F-91405, Orsay, France}

\begin{abstract}
The role of activity on the hydrodynamic dispersion of bacteria in a model porous medium is studied by tracking thousands of bacteria in a microfluidic chip containing randomly placed pillars. 
We first evaluate the spreading dynamics of two populations of motile and non-motile bacteria injected at different flow rates. In both cases, we observe that the mean and the variance of the distances covered by the bacteria vary linearly with time and flow velocity, a result qualitatively consistent with the standard geometric dispersion picture. However, quantitatively, the motiles bacteria display a systematic retardation effect when compared to the non-motile ones. Furthermore, the shape of the traveled distance distribution in the flow direction differs significantly for both the motile and the non-motile strain, hence probing a markedly different exploration process. For the non-motile bacteria, the distribution is Gaussian whereas for the motile ones, the distribution displays a positive skewness and spreads exponentially downstream akin to a Gamma distribution. The detailed microscopic study of the trajectories reveals two salient effects characterizing the exploration process of motile bacteria : (i) The emergence of an  “active” retention effect due to  an extended exploration of the pore surfaces, (ii) an enhanced spreading at the forefront due to the transport of bacteria along ``fast-tracks'' where they acquire a velocity larger than the local flow velocity. We finally discuss the practical applications of these effects on the large-scale macroscopic transfer and contamination processes caused by microbes in natural environments.

\end{abstract}

\maketitle
Understanding the transport of micro-organisms in heterogeneous media is a question dealing with a large variety of scientific and technological domains such as bacteriology, ecology, sciences for environment, petroleum research or medicine. For example, bacteria are now used as vectors for fighting cancers \cite{Anderson2006, Felfoul2016}. The oil industry also considers the potential of bacteria injection to enhance oil recovery \cite{Brown2010}. Nowadays, bio-remediation techniques are developed in which contaminants trapped in the ground, are targeted then decomposed or fixed by bacteria \cite{BookRemediation}. Some biotechnologies require the isolation of specific microbial strains and these processes need adapted filtration or sorting techniques \cite{Karimi2013}. 

A better understanding of how microorganisms are transported trapped or dispersed in disordered porous media is not only a key to the development of future innovative applications but also may shed new lights on the strategies that micro-organisms use to maximize their survival and proliferation abilities in natural conditions.

In the past decades, modeling the transport of micro-organisms was essentially driven by an increasing concern about pollution of ground waters \cite{Ginn2002, Pandey2014}. The approaches used to model the transport in porous media are currently based on the standard advection-dispersion equation including biological processes such as growth and mass exchange with the grains, through phenomenological coefficients derived from breakthrough curves \cite{Corapcioglu1985, Peterson1989, Hendry1999, Foppen2005, Tufenkji2007, Bai2016}. The focus has been mostly put on the adhesion properties influenced by chemical processes like pH or ionic strength. The final outcome of these analyzes is often disappointing as most of the reports conclude on difficulties to scale up laboratory column experiments \cite{Lutterodt2009, Lutterodt2011}. Surprisingly, little is known about the influence of microbial motility on the retention processes inside the pores or the way the swimming ability contributes to the transport in confined channels although some studies bring the evidence of an undeniable influence \cite{Camesano1998, Becker2003}.

Recent developments in microfluidic techniques provide a new and efficient tool that can be used to visualize transport processes of bacteria and allow to assess the influence of well- controlled environments. For example, Ford and co-workers \cite{Tao2009} quantified the enhancement of transverse migration of bacteria due to chemotaxis. Additionally, other developments in microfluidic technology provided a way to observe the movement of bacteria in ``simple" flows and to study the coupling between bacteria motility and flows in channels \cite{Rusconi2014}, in constrictions \cite{Altshuler2013}, close to surfaces \cite{Hill2007, Kaya2012, Marcos2012} at stagnation points \cite{Mino2018} or in corners \cite{Morales2015}. 

Our study demonstrates how physical processes associated to the interplay between motility and flow in a disordered porous medium, lead to significant effects on the macroscopic transport properties. To this purpose, we designed a microfluidic environment in which motility and pore geometry are the dominant ingredients influencing the hydrodynamic dispersion of a bacterial fluid. The microfluidic channel includes some of the random structural heterogeneities of natural pore structures. We also used motile and non-motile bacteria that do not stick to the surfaces and performed in a large window of flow rates a study that provides important clues to help developing new physical transport models. 

\section{Material and methods}
 
\subsection{Microfluidic chip fabrication}
The microfluidic chip is designed by standard photo-lithography techniques. The channel geometry is a serpentine of rectangular section (width = 500~$\mu$m, height = 100~$\mu$m) filled with circular pillars (see Fig. \ref{fig:fig1}b). The pillar centers are randomly distributed along the channel and their diameter are randomly chosen among four values (20, 30, 40 and 50~$\mu$m). Channels are designed to be filled at $33\%$. The average diameter of the obstacles is thus $d$ = 35~$\mu$m, the average distance between two close neighbors is $\sim$10~$\mu$m. Channels are made of polydimethyle siloxane (PDMS) and the chip is bonded onto a glass plate (5.5~cm $\times$ 5.5~cm) using a plasma cleaner. PMDS is permeable to oxygen allowing a continuous oxygenation of the suspension ensuring the constant swimming activity of the bacteria during the experiments \cite{Douarche2009}.
 
\subsection{Strain and culture}
A fluorescent \textit{Escherichia coli} (\textit{E. coli}) RP437 strain is used. The fluorescence is obtained by transforming the wild type \textit{E. coli} RP437 with a plasmid coding for a yellow fluorescent protein (YFP). The bacteria are cultured overnight at 30$^\circ$C and shaken at 240 rpm in M9 minimal medium supplemented with 1 mg/ml casamino acids, 4 mg/ml glucose and a 25$~\mu$g/ml chloramphenicol at 0.05$\%$. The growth medium is then removed by centrifugation (2300~g for 10~min). Bacteria suspension is then rinsed with milliQ water to remove any residuals from the overnight culture. The bacterial population is finally resuspended into a motility buffer containing 10~mM phosphate buffer (pH=7), 0.1~mM EDTA, 1$\mu$M L-methionine, 10~mM sodium L-lactate. To avoid bacterial sedimentation, the suspension is mixed with Percoll (1 v/v). In this suspending medium, bacteria are able to live and swim but do not divide. The optical density (OD) is measured at 600~nm using a Spectrophotometer BEL SP 1105 in order to determine the bacterial concentration. All experiments are performed at a low cell concentration (OD $\sim2.5\times10^{-3} \sim$3~bact/$\mu$l) but high enough to get good statistics for tracking measurements. Under those conditions, the average swimming velocity is measured to be $\overline{v_b}=19~\pm 4~\mu$m.s$^{-1}$ and remains roughly constant for several hours. Non-motile bacteria are prepared by keeping the suspension at $4^{\circ}$C for $5$ hours. The suspension is then maintained at 22$^{\circ}$C for 30~min before its use. No sign of motility is  detected over the entire duration of the experiments (\textit{i.e.} $\sim$4 hours). 

\subsection{Experimental protocol}
\begin{figure}
\begin{center}
\includegraphics[width=8.5cm]{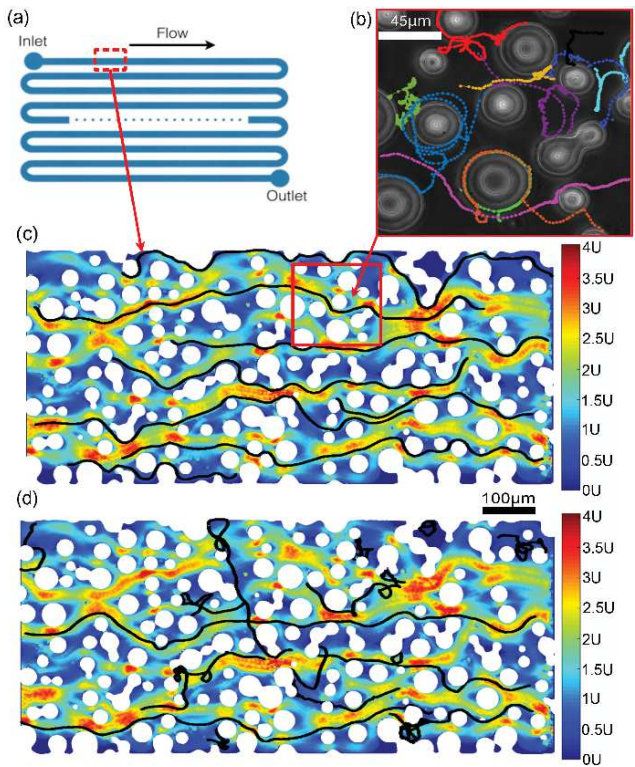}
\caption{(a) Schematic view of the overall microfluidic chip. (b) Colored lines representing trajectories of motile bacteria in the porous space. (c) and (d) Superimpositions of the velocity field obtained with passive tracers and \textit{E. coli} trajectories (black solid lines) for non-motile (c) and motile (d) bacteria during $\Delta t$ = 10~s at a flow rate $U$ = 36~$\mu$m.s$^{-1}$. White circles represent the pillars (average diameter $d$ = 35~$\mu$m). Images are recorded with a 10$\times$ objective. The field of view is 1~mm $\times$ 0.5~mm.}
\label{fig:fig1}
\end{center}
\end{figure}

Observations are performed with a camera mounted on a Leica DMI 6000 inverted microscope driven by $\mu$Manager software \cite{Edelstein2014}. Two cameras are used: (i) a IDS CMOS camera for phase-contrast visualization and (ii) a Hamamatsu ORCA-Flash4.0 camera when the fluorescence mode is preferred. Most of the movies are recorded at different places located at about 2 cm from the inlet of the channel which corresponds to a distance of approximately 20 000 cell bodies or 600 grain sizes. First, the microfluidic device is filled with motility buffer and percoll (50\%). A 100~$\mu$l syringe is filled with the bacterial suspension and loaded on a computer-controlled Nemesys pump. Most of the experiments are done with a 10$\times$ objective the field of depth of which is $\sim$30~$\mu$m (about 1/3 of the channel height). Before the image acquisition, the objective is moved vertically to focus on the middle of the channel. The images of Fig. \ref{fig:fig1}b and the movie are obtained using a 63$\times$ objective (field of depth 1~$\mu$m). Sequences of $3000$ images are recorded with a frame rate that depends on the flow velocity. It is chosen such that the average distance traveled by the bacteria between two successive frames is $\sim$5~$\mu$m. Before the acquisition of the image sequence, 3 to 5 short sequences of 100 images are taken every 10 min and processed systematically to verify the steadiness of the flow. 

\subsection{Image analysis and bacteria tracking}
Bacteria tracking is done using the ``trackmate" plug-in of the image analysis free-ware Fiji \cite{Fiji}. Prior to the analysis, the average image obtained over the whole sequence lasting between 32 and 450~s is subtracted from each image. The tracking procedure has two steps: (i) bacteria detection and (ii) frame by frame linking of positions to built the individual tracks. The output is a datafile giving for each detected bacterium its position $x \vec{e}_x + y \vec{e}_y$ as a function of time $t$. The Figs. \ref{fig:fig1}c and \ref{fig:fig1}d show examples of trajectories obtained with this method. Thousands of trajectories are obtained at the same field of view along the channel for motile and non-motile bacteria but also for latex beads of diameter 2~$\mu$m. The flow velocity field $\vec{v}(x,y)$ is derived from these passive tracer trajectories. The coarse-grained spatial resolution used to determine the flow field corresponds to a square of 5~px~$\times$~5~px (2.75~$\mu$m $\times$ 2.75~$\mu$m). Figs. \ref{fig:fig1}c and \ref{fig:fig1}d show a typical velocity field obtained using this method and superimposed with a few trajectories of non-motile (Fig. \ref{fig:fig1}c) and motile (Fig. \ref{fig:fig1}d) bacteria. The conversion between the flow rate $Q$ imposed by the pump and the average flow velocity $U$ is obtained from a series of calibration experiments performed with the passive tracers.

\section{Results}
 
Differences between motile and non-motile bacteria are qualitatively illustrated in the trajectories displayed on Figs. \ref{fig:fig1}c and \ref{fig:fig1}d. While the trajectories of non-motile bacteria are primarily oriented along the flow direction and follow the streamlines (Fig. \ref{fig:fig1}c), the motile bacteria display more erratic trajectories with a significant deviation from the flow lines (Fig. \ref{fig:fig1}d). Magnification at the level of the obstacles (Fig. \ref{fig:fig1}b) reveals that the motile bacteria trajectories are interspersed with moments in which the bacteria change direction, move upstream, and travel back and forth from the vicinity of one obstacle to another, much like a ball in a pinball game.

\begin{figure*}
\begin{center}
\includegraphics[width=\textwidth]{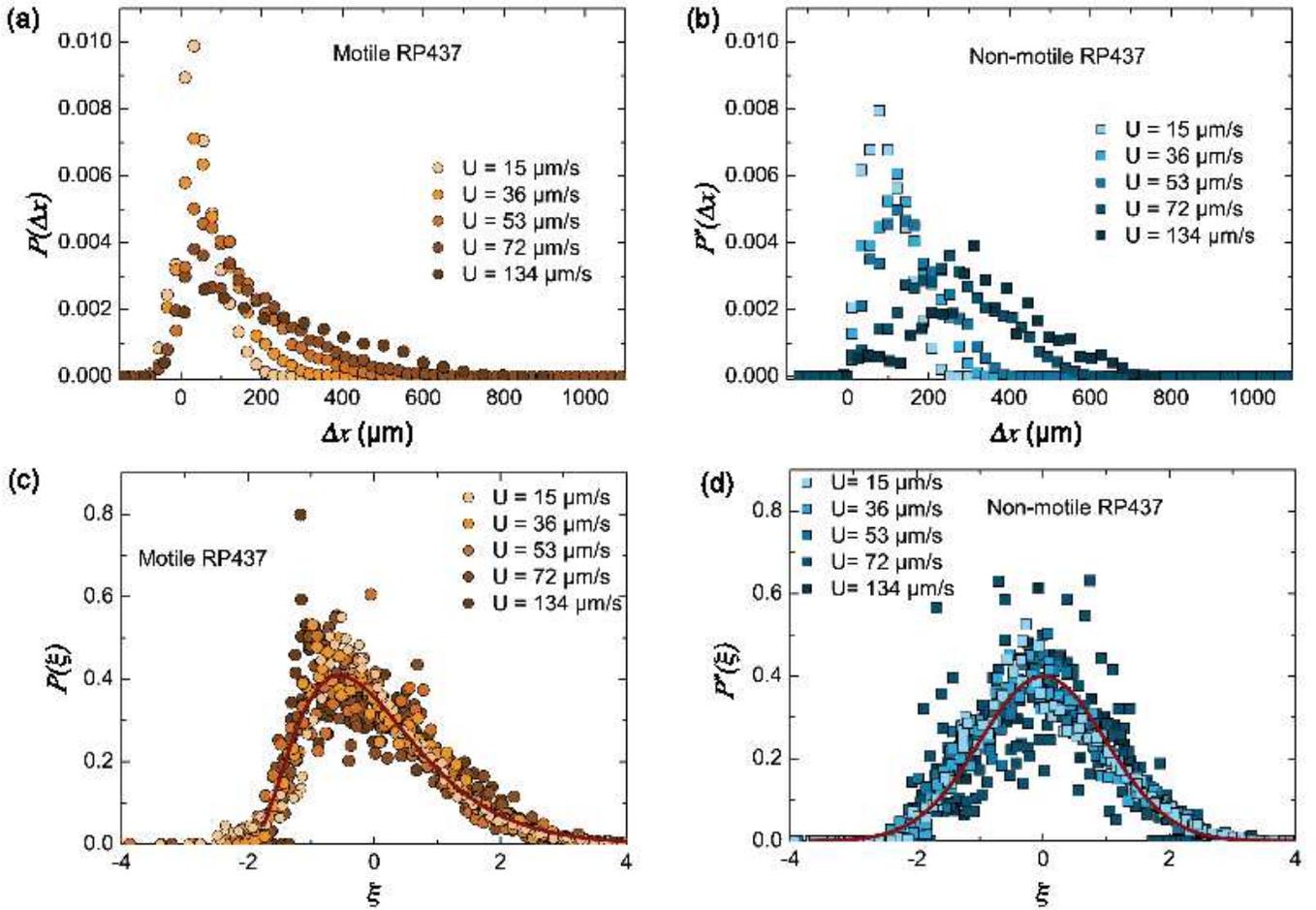}
\caption{(a) and (c) Distributions $P(\Delta x,\Delta t)$ of the distances $\Delta x$, traveled along the flow direction at a time $\Delta t = 3$~s, for motile (a) and non-motile (c) bacteria at different flow velocities. (b) and (d) Normalized distributions $P^*(\xi)$ where $\xi=\frac{\Delta x -\overline{\Delta x}}{\sigma_x}$ for the motile (b) and the non-motile (d) bacteria. For each flow velocity, five different $\Delta t$ are represented ($\Delta t =$ 1, 3, 5, 7 and 9~s). Solid red line: (b) a Gaussian distribution of zero mean and unit variance and (d) a best fit from a $\Gamma$ distribution.}
\label{fig:fig2}
\end{center}
\end{figure*}

Quantitatively,  we characterize the transport and dispersion properties by computing the distributions of distances traveled by the bacteria both along the flow direction ($\Delta x$) and the transverse direction ($\Delta y$) over a time interval $\Delta t$. Fig. \ref{fig:fig2} shows the distributions $P(\Delta x,\Delta t)$ for a fixed time interval $\Delta t=3$~s  for non-motile (Fig. \ref {fig:fig2}a) and motile (Fig. \ref {fig:fig2}b) bacteria, at different flow rates. Significant differences in the spreading dynamics and in the shape are immediately visible. In the following, we provide an account for the transport of motile and non-motile bacteria by a quantitative analysis of these distributions.  We define the averages of the displacement distributions over the trajectories ($\overline{\Delta x}$ and $\overline{\Delta y}$) as well as the standard deviations ($\sigma_{x}$  and $\sigma_{y}$)  along the $x$ and $y$ axis respectively. We also determine the distributions of the adimensionalized variables : $\xi= \frac{\Delta x-\overline{\Delta x}}{\sigma_{x}}$ and the skewness of the distribution along the flow $Sk_x$.

\subsection{Dispersion processes}
To highlight the differences between the motile and the non-motile bacteria, we characterize the shapes of the distributions of the adimensionalized transport distances along the flow ${P}^*(\xi)$ for the motile (see Fig. \ref{fig:fig2}b) and for the non-motile (see Fig. \ref{fig:fig2}d) bacteria.

Importantly these distributions collapse on the same curve for all flow velocities $U$ and for time intervals corresponding to a mean displacement larger than one and up to 6 pore sizes ($\sim$~210~$\mu$m) as long as the measurement can be performed. Then it shows that above the pore distance, one gets a converged stochastic process characterizing the transport and the dispersion of bacteria in the porous medium. For the non-motile bacteria, the distribution is very close to a normal distribution (a Gaussian of zero mean, unit variance and zero skewness). This result is the one expected for classical geometrical dispersion of inert species in porous flows \cite{Saffman1959, Bear1971}. For motile bacteria, however, the rescaled distribution displays a significant mean skewness ($Sk_x \sim 0.85$) characterizing both a forward spreading and a retardation effect that can be seen on the maximum position (the mode) standing at a negative value ($\sim -0.5 $). For $\xi_s>-2$, the distribution can be approximately fitted using a Gamma distribution which highlights the exponential decay of the forefront: ${P}^*(\xi) = \Gamma(k)^{-1} \xi_0 ^{-k} (\xi-\xi_s) ^{k-1} exp(- (\xi-\xi_s) / \xi_0 )$ with $k \sim 3.366$, $\xi_s = -2$ and $\xi_0 \sim 0.612$. 
Therefore at the front,  the rescaled distribution decays more slowly for motile than for non-motile bacteria which is the sign of an enhanced transport process over a mesoscopic scale larger than the pore size.\\

\subsection{Average transport properties}
\begin{figure}
\includegraphics[width=8.5cm]{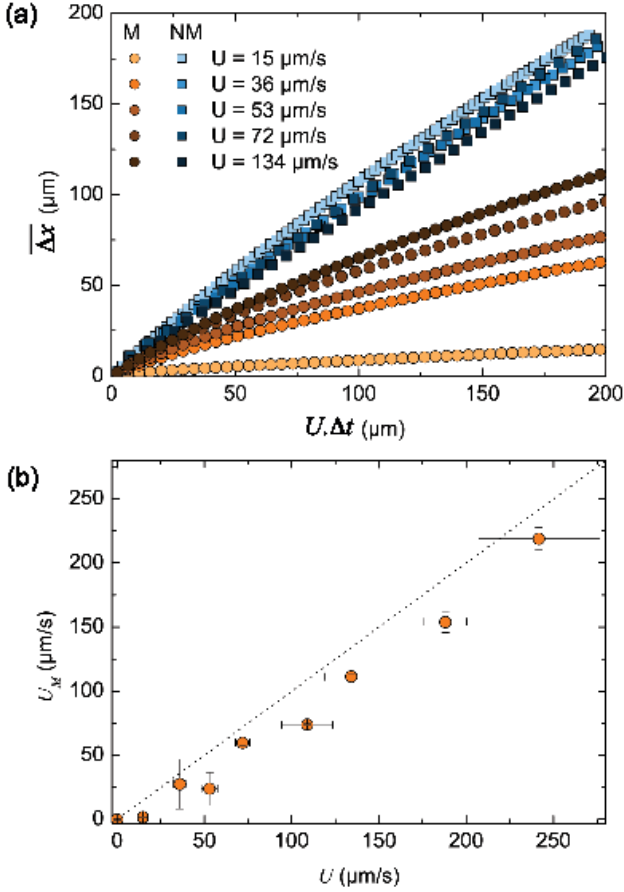}
\caption{(a) Mean positions of the bacteria population $\overline{\Delta x}$, as a function of the mean distance traveled by the fluid $U.\Delta t$, both for motile (M) and non-motile (NM) bacteria and for different flow rates. (b) Mean transport velocity of the motile bacteria $U_M$ as a function of the mean flow velocity $U$. The dotted line represents the $y=x$ function.}
\label{fig:fig3}
\end{figure}

We study the influence of the flow velocity on the average position of the bacterial population. For all flow rates and for different time intervals $\Delta t$, the mean displacements $\overline{\Delta x}$  are computed. Fig. \ref{fig:fig3}a displays the average displacements as a function of the average distance $U.\Delta t$ traveled by the fluid during the same time. We see that all data obtained for the non-motile bacteria (square symbols in Fig. \ref{fig:fig3}a) roughly collapse onto a line of slope~$\sim 1$. This result indicates that the non-motile bacteria progress in the porous medium with the average velocity of the fluid as would do passive tracers. However for motile bacteria, we identify significant differences (circle symbols in Fig. \ref{fig:fig3}a). For a given mean flow velocity $U$, the average distance $\Delta x$ also increases linearly with time but with a slope smaller than $1$. This slope depends on the mean flow rate $U$ hence probing a retardation effect due to motility. A linear fit of the data gives an estimate of the average transport velocity $U_{M}$ of the motile bacteria. On Fig. \ref{fig:fig3}b, $U_M$ is plotted as a function of $U$ and we see for $U>50$~$\mu$m.s$^{-1}$ an offset of the order of $20\pm 12~\mu$m.s$^{-1}$.

\subsection{Dispersion dynamics}
\begin{figure}
\includegraphics[width=8.5cm]{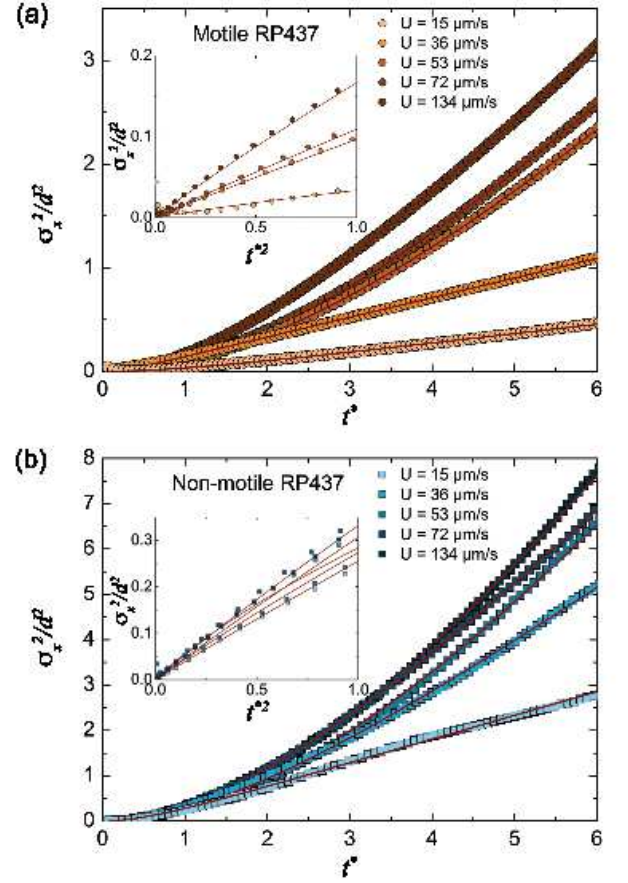}
\caption{(a) and(b) Variation of the dimensionless quadratic displacement $\dfrac{\sigma^2_{x}}{d^2}$ as a function of the normalized time ${t^*}=\frac{U.\Delta t}{d}$ for different average flow velocities respectively for motile (a) and non-motile (b) {\it E. coli}. Solid lines: adjustments by the function $f({t^*})=\alpha (\beta {t^*}-(1-e^{-\beta {t^*}}))$. Insets represent the dimensionless quadratic displacement as a function of $t^{*2}$ for less than one pore size.}
\label{fig:fig4}
\end{figure}

To quantify the spreading of the bacteria, we display the quadratic displacement of the longitudinal distance traveled by the bacteria rescaled by the mean size of the obstacles $\dfrac{\sigma^2_{x}}{d^2}$ as a function of the rescaled time ${t}^*= \dfrac{U.\Delta t}{d}$ (see Fig. \ref{fig:fig4}) for different average flow velocities. The data obtained for the motile (Fig. \ref{fig:fig4}a) and the non-motile (Fig. \ref{fig:fig4}b) bacteria, show essentially the same behavior. At short periods of time, a ballistic regime where $\sigma^2_{x}$ varies quadratically with time and at long periods of time (see insets in Fig. \ref{fig:fig4}), a transition towards a diffusive regime for which $\sigma^2_{x} \sim t^*$. To extract the effective longitudinal diffusion coefficient, we use the standard Fürth function : $f({t^*})=\alpha (\beta {t^*}-(1-e^{-\beta {t^*}}))$ derived from an Ornstein-Uhlenbeck process \cite{Uhlenbeck1930, Wu2006} using $\alpha$ and $\beta$ as fitting parameters. The solid lines in Figs. \ref{fig:fig4}a and \ref{fig:fig4}b show the  adjustments by a least square regression of our data. The adjustment is very good for all the flow velocities used and up to ${t^*} \sim 6$ pore sizes. The fitting parameters are then used to determine the dispersion coefficients $D_L=\dfrac{\alpha \beta d U}{2}$. The confidence intervals of $\alpha$ and $\beta$ were used to calculate the error bars of the dispersion coefficients. The cross-over times correspond in all cases to a traveling distance comparable to a pore size. 

\begin{figure}
\includegraphics[width=8.5cm]{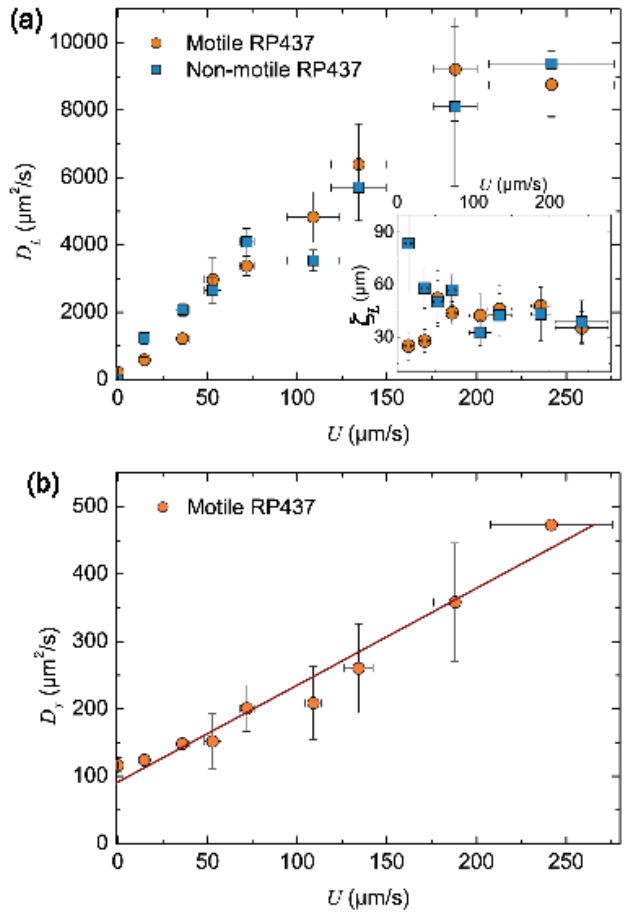}
\caption{(a) Longitudinal dispersion coefficient $D_L$ as a function of $U$ for motile and non-motile bacteria. Inset: Dispersivity $\zeta_L^{NM} = \dfrac{D_L^{NM}}{U}$ and $\zeta_L^{M} = \dfrac{D_L^M-D_0^M}{U}$ as a function of $U$ for non-motile and motile bacteria repsectively. The dotted line represents the constant function $y =$~50 $\mu$m. (b) Transverse dispersion coefficient $D_T$ of motile bacteria. The solid line represents a linear regression of the data yielding $\zeta_T^{M}=1.5 \pm 0.1$~$\mu$m.}
\label{fig:fig5}
\end{figure}

The dispersion coefficients for non-motile bacteria ($D^{NM}_{L}$) and motile bacteria ($D^{M}_{L}$) are displayed on Fig. \ref{fig:fig5} as a function of the mean flow velocity $U$. For non-motile bacteria the dispersion coefficient is independent of the flow rate and we extract a dispersivity length scale $\zeta^{NM}$ such that  $D^{NM}_{L} = \zeta^{NM}_{L} U$. One finds $\zeta^{NM}_{L}= 50~(\pm 16)$~$\mu$m, on the order of the pore size which is a standard feature of geometrical dispersions occurring in disordered porous media \cite{Hendry1999, Bai2016}.

For motile bacteria, we extract an effective dispersivity length scale $\zeta^{M}$ using the relation  $D^{M}_{L} = D^{M}_0+ \zeta^{M}_{L} U $ where $D^{M}_0 = 210$~$\mu$m$^2$.s$^{-1}$ is the zero-flow diffusivity extracted from the experiments. Note this value is three orders of magnitude larger than its equivalent for non-motile bacteria. Therefore in the experimental uncertainties, the value of $\zeta^{M}$ seems to be equivalent to $\zeta^{NM}$ (see inset of Fig \ref{fig:fig5}a).

We also find that bacteria motility induces a significant enhancement of the transverse exploration process. From a similar analysis as for longitudinal dispersion, we extract a transverse dispersion coefficient $D^{M}_{T}$  of the motile bacteria (for the non-motile ones, the effect is quite undetectable). The values of $D^{M}_{T}$ are represented in Fig. \ref{fig:fig5}b and the linear regression of the data yields: $D^{M}_{T} = D^{M}_0+\zeta^{M}_{T} U $ with a dispersivity $\zeta_T^{M}=1.5 \pm 0.1$~$\mu$m much smaller than the longitudinal one. This transverse dispersion is yet another feature that distinguishes the transport of motile microorganisms from non-motile ones.

The transport characteristics obtained for motile and non-motile bacteria reveal that motility has two major effects that may seem contradictory and  contra-intuitive : (i) a retardation of the transport for a large number of motile bacteria and at the same time, (ii) a rapid downstream progression of some others. To help understand these two antagonistic observations, we analyze in the following the statistical properties of the trajectories from a more microscopic point of view in order to  identify some elementary mechanisms influencing the transport process.

\subsection{Dynamical trapping}
\begin{figure}
\includegraphics[width=8.5cm]{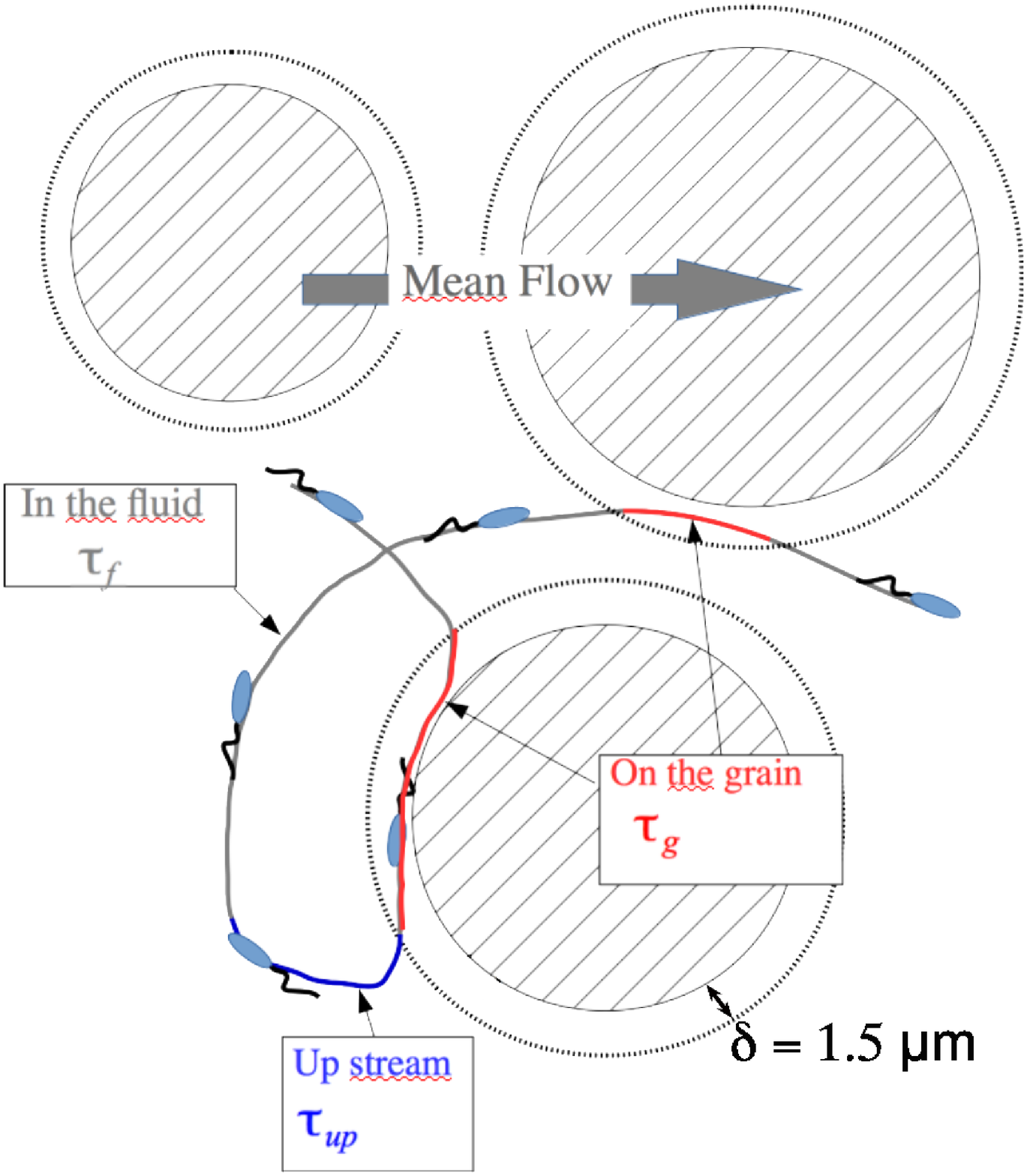}
\caption{Schematics of a bacterium trajectory and the corresponding parameters as defined in the text. $\tau_{up}$ is the period of time spent with a negative $v_x$, $\tau_g$ is the period of time spent on the grain vicinity \textit{i.e.} when the distance bacteria/obstacle is smaller than $\delta$ = 1.5~$\mu$m and $\tau_f$ is the period of time spent in the fluid when the distance bacteria/obstacle is larger than $\delta$ = 1.5~$\mu$m.}
\label{fig:fig6}
\end{figure}

\paragraph*{Staying in the grain vicinity}
First, we study the influence of the motility on the presence of bacteria in the vicinity of the pillar surfaces (the grains). The trajectories are divided into ``fluid" and ``grain" sections (see Fig. \ref{fig:fig6}) characterizing the presence of the microorganisms in these domains. 

\begin{figure}
\includegraphics[width=8.5cm]{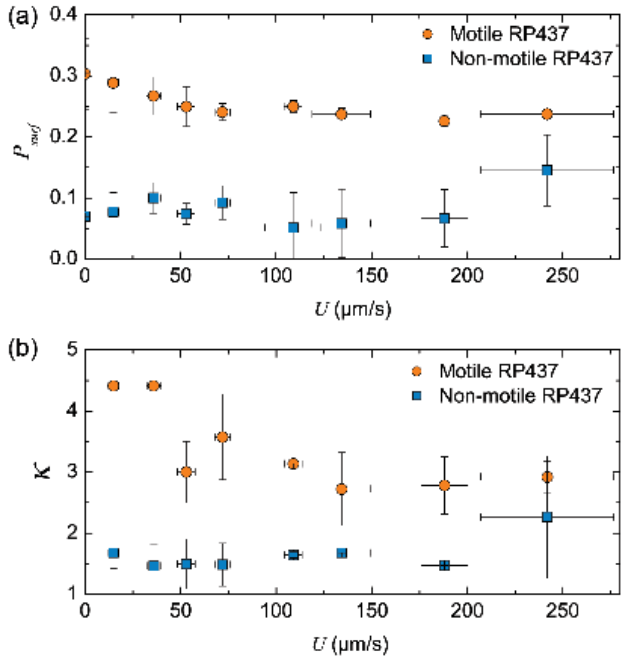}
\caption{(a) Average time spent by the bacteria swimming in the vicinity of an obstacle $\tau_{g}$ normalized by the total time $\tau_g+\tau_f$ where $\tau_f$ is the average time spent in the fluid as a function of $U$. (b) Mean number of occurrences a bacterium gets in contact with an obstacle per unit of time $\tau=d/U$.}
\label{fig:fig7}
\end{figure}

The ``fluid" sections correspond to sub-parts of the trajectories where the distance bacteria/obstacle is larger than $\delta = 1.5$~$\mu$m while the ``grain" sections correspond to the parts of the trajectories for which the distance bacteria/obstacle is less than 1.5~$\mu$m. The average duration of the ``fluid" segments, $\tau_f$, and ``grain" segments, $\tau_g$, are computed and averaged over all the trajectories. The relative fractions of time, $P_{surf}=\dfrac{\tau_g}{\tau_f + \tau_g}$, for the motile bacteria and the non-motile bacteria are shown in Fig.\ref{fig:fig7}a. One observes a striking difference between the motile and the non-motile bacteria. 

For non-motile, the average flow velocity $U$ has no influence on $P_{surf}$ and its value remains significantly lower than the value of the motile ones. The relative amount of time spent around the obstacles is approximately 4 times greater for the motile bacteria than for the non-motile ones and this effect seems to be more pronounced when flow velocities become comparable to the bacteria velocity.

We also quantify $\kappa$, the frequency of encounters of the bacteria with the grain surfaces. Then we compute the number of contacts divided by the track duration and average this quantity over all tracks. On Fig. \ref{fig:fig7}b we show the values of $\kappa$ for motile and non-motile bacteria. For non-motile bacteria this frequency remains constant over the whole range of fluid velocities but also remains significantly lower than its counterpart for motile bacteria. The ratio can be as large as 5 when the fluid velocities become comparable to the bacteria swimming velocity. 

These sharp differences prove that motility promotes the flow of bacteria towards the obstacles and at the same time, increases the residence time of the swimmers at the surfaces.

\begin{figure}
\includegraphics[width=8.5cm]{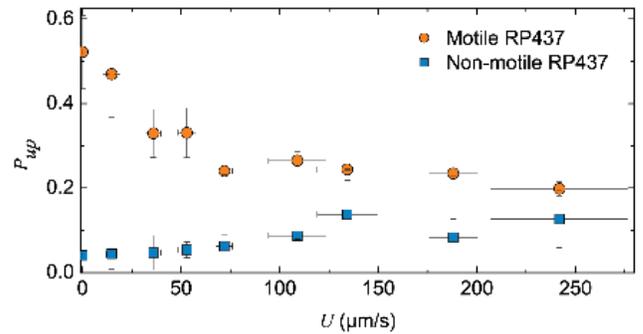}
\caption{Average duration of the upstream displacement $\tau_{up}$ normalized by the total time $\tau_{up}+\tau_{down}$ where $\tau_{down}$ is the average duration of the downstream ($v_x>0$) displacement as a function of $U$.}
\label{fig:fig8}
\end{figure}

\paragraph*{Upstream motion}
Upon careful observation of movies recorded at high magnification (see Movie in the SM), we see an impressive number of motile bacteria moving upstream (\textit{i.e.} with a negative velocity in the laboratory reference frame), while this behavior seems absent for non-motile bacteria. To quantify the upstream swimming, trajectories are segmented into periods during which the bacteria either move downstream or upstream. The average durations of these portions of trajectories, $\tau_{up}$ and $\tau_{down}$, are calculated. Fig. \ref{fig:fig8}  displays the relative time $P_{up} = \dfrac{\tau_{up}}{\tau_{up} + \tau_{down}}$ showing a clear difference between the values measured for motile and non-motile bacteria: the periods during which motile bacteria move upstream are very long whereas for non-motile ones the upstream motion is almost absent.

This analysis clearly demonstrates that motility favors the flow of bacteria toward regions of low velocity, most of the time close to the grains where they perform upstream motion. This is a dynamical trapping effect that will contribute to a global transfer of motile bacteria at an average velocity lower than the fluid velocity. This
observation is a microscopic explanation of the retardation effect reported previously for the dispersion curves.

\subsection{Rapid downstream migration}
\begin{figure}
\includegraphics[width=8.5cm]{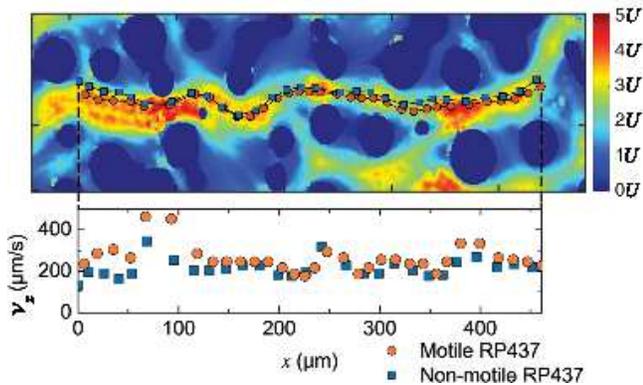}
\caption{Overlap of two close trajectories of a motile (orange circles) and a non-motile (blue squares) bacteria with the local fluid velocity. The inset shows the speed of the bacteria $v_x$ as a function of the longitudinal position $x$. Experiments are performed at $U =$ 72~$\mu$m.s$^{-1}$.}
\label{fig:fig9}
\end{figure}

Let us come back now to the second observation concerning the rapid downstream progression of some motile bacteria. A close look at the velocity field reveals the existence of channels connecting the inlet and the outlet of the porous medium (within the visualization window) and along which the fluid velocity is at least twice the average fluid velocity $U$ (see Fig. \ref{fig:fig9}). We called these channels the ``fast tracks''. The bacteria moving at a stream-wise velocity at least twice the mean flow velocity will be called \textit{fast} (or \textit{slow} otherwise). To illustrate this ``fast-track" mechanism enhancing the transport of motile bacteria in the fastest channels, we overlay onto the velocity map, two very close trajectories, one for a motile bacterium and the other for a non-motile one (see Fig. \ref{fig:fig9}). 

In the inset, the velocities along the stream-wise coordinate are displayed. We see in this example,  that the motile bacterium is always traveling faster than the non-motile one. We also notice a systematic increase in the velocity difference as they cross the pore constrictions. The flow in these regions seems to align the bacteria along the flow direction thus enhancing their mean transport velocity. Interestingly, the effect of stabilization by constriction was also described and discussed recently by Potomkin \textit{et al.} \cite{Potomkin2017}.

In Fig. \ref{fig:fig9}a, we display the mean velocity of the fast bacteria $V_{fast}$ as a function of the mean flow $U$. Then, it appears that the population of fast motile bacteria is transported at a velocity in average higher than the population of fast non-motile ones. We anticipate that the fast bacteria are essentially transported along the ``fast-tracks''.
\begin{figure}
\includegraphics[width=8.5cm]{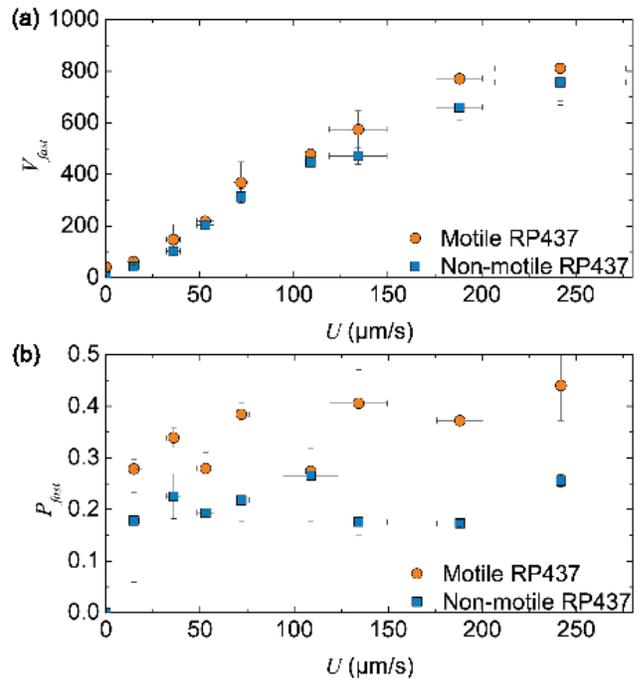}\\
\caption{(a) Speed of the fastest detected bacteria as a function of the mean flow velocity $U$. (b) Average duration, $\tau_{fast}$, of the sequences during which a bacteria is moving with a speed $v_x$ larger than twice the average flow velocity, normalized by $\tau_{fast}+\tau_{slow}$ where $\tau_{slow}$ is the averaged duration of the sequences during which $v_x<2U$.}
\label{fig:fig10}
\end{figure}

To further quantify this effect, each trajectory is split into ``fast" and ``slow" sections. The relative duration of the ``fast" segments $P_{fast}=\dfrac{\tau_{fast}}{\tau_{fast}+\tau_{slow}}$, are displayed in Fig. \ref{fig:fig9}b for both motile and non-motile bacteria and for all the flow velocities $U$ studied. 
It is immediately noted that - except for the lowest velocities - motile bacteria stay in the fast streams longer than the non-motile bacteria.    

\section{Summary and conclusion}  
To conclude, we have shown that microbial motility has a significant impact on the average transport and dispersion properties of bacterial suspensions flowing inside a porous medium.

By tracking swimming and non-swimming bacteria up to a transport scales of typically 6 times the pore size, we identify a phenomenology contrasting sharply between the motile and the non-motile bacteria. For the non-motile, one recovers the classical results of geometrical hydrodynamic dispersion characterizing the transport of inert species inside a porous medium \cite{Saffman1959, Bear1971}. This is essentially the phenomenology expected for dilute colloidal suspensions with no adhesion on solid surfaces, provided a particle size smaller than the pore throat. In this case the mean transport velocity is the mean flow velocity and the longitudinal dispersion coefficient is proportional to the flow velocity, with a dispersivity length-scale congruent to a pore size.  
 
However for the active species the ability to swim across the flow lines and eventually to dwell at surfaces or, on the opposite, to take preferentially fast channels, changes drastically the distribution of bacteria in the flow.  The shape of the mean distribution of longitudinal displacements at a given  lag-time, is no longer a Gaussian distribution as for the non-motile ones. In the stream-wise direction, the mode of the distribution comes before the mean hence characterizing a retardation effect due to the motile character of the bacteria. The curve is akin to a Gamma distribution characterized by a positive skewness and an up-stream front decaying exponentially. Remarkably, this non-Gaussian distribution seems to hold for mean transport distances significantly larger than the pore size. For motile bacteria the mean transport velocity is systematically retarded with respect to the non-motile ones. This "dynamical trapping" effect as we describe it, does not resulting from any chemical or physical interaction often invoked to explain transport retardation, but essentially borne in the swimming activity of the bacteria spending a significant time close to the surfaces and also swimming upstream. This is in stark contrast with the non-motile species which essentially follow the flow lines. On the opposite, we also measured a contribution that could explain the exponential forefront as the motile bacteria seem to take fast channels for a significantly long time and add up their swimming contribution to the maximal flows.  Interestingly, this ``fast-track'' effect identified in the context of a model porous medium, seems to be consistent with the observations of several laboratory-scale columns and field experiments who reported the presence of early breakthroughs of microorganisms \cite{Hornberger1992, McCaulou1994, Stumpp2011}. 

In spite of these marked differences between motile and non-motile species, the scaling of the mean longitudinal dispersion with the flow velocity are qualitatively similar (proportional to the mean flow). Importantly, the motile bacteria display transverse dispersion, an effect which is almost undetectable for the non-motile. 

From these results, emerges a new understanding on how the motility of many micro-organisms influences in depth the transport processes and the spatial distribution in natural environments such as porous soils, fractured rocks or even biological networks. We already foresee two important practical implications. First, for bacterial communities transported in a flow, the active retention effect leads to a thorough exploration of surfaces. This would influence strongly the chances for adhesion or biofilm formation. Second, because motility favors an efficient longitudinal transport for a sub-population of swimmers, it will increase the volume explored by pioneering bacteria with important implications on the development of forefront contamination and colonization.

Remarkably, in the last years, active matter studies have brought to the front many paradigmatic shifts in the understanding of the out-of-equilibrium organization of motile bacterial suspensions \cite{Marchetti2013}. In this context, standard notions such as the equations of state  \cite{Solon2015}, phase transitions \cite{Ginelli2010, Gregoire2004}, Brownian of Fickean diffusion  \cite{Wu2000, Galajda2007} or the rheological response \cite{Hatwalne2004, Sokolov2009, Rafai2010, Saintillan2010, Lopez2015} had to be deeply revisited to account for the crucial importance of motility. Consequently, transport, dispersion or filtration of motile micro-organisms in porous media, is likely to yield results that differ qualitatively from those of passive colloids.

These effects are novel transport properties in addition to the many singular features already identified for active matter. But it is also interesting to realize that they impact positively the natural ability of bacterial populations to survive, grow and reach their ecological niche.

\section*{Acknowledgements}
This work is supported by public grants overseen by the French National Research Agency (ANR): (i) ANR Bacflow AAPG 2015 and (ii) from the ``Laboratoire d'Excellence Physics Atom Light Mater" (LabEx PALM) as part of the ``Investissements d'Avenir" program (reference: ANR-10-LABX-0039). We acknowledge support by Universidad de Buenos Aires (UBACyT No.20020130100570BA) and the LIA PMF-FMF (Franco-Argentinian International Associated Laboratory in the Physics and Mechanics of Fluids). H.A. and A.C. thank D. Bouville for his help in the clean room.


\begin{thebibliography}{}
\bibitem{Anderson2006} J.C. Anderson, E.J. Clarke, A.P. Arkin and C.A. Voig, J. Mol. Biol. {\bf{355}}, 619 (2006); 
\bibitem{Felfoul2016} O. Felfoul et al., Nature Nanotech. {\bf{11}}, 941--947 (2016).
\bibitem{Brown2010} L. R Brown, Curr. Opin.  Microbiology {\bf{13}}, 316  (2010)
\bibitem{BookRemediation} Advanced Groundwater Remediation: Active and Passive Technologies, ed. F.-G. Simon,T. Meggyes, C. McDonald (ThomasTelford publishing, London, 2002).
\bibitem{Karimi2013} A. Karimi, S. Yazdi and A. M. Ardekani, Biomicrofluidics {\bf{7}}, 021501 (2013).
\bibitem{Ginn2002} T.R. Ginn et al. Adv. Water Resour. {\bf{25}}, 1017 (2002).
\bibitem{Pandey2014} P.K. Pandey et al. AMB Express {\bf{4}}, 51 (2014).
\bibitem{Corapcioglu1985} M.Y. Corapcioglu and A. Haridas, Adv. Water Resour. {\bf{8}}, 188 (1985).
\bibitem{Peterson1989} T.C. Peterson and R.C. Ward,  JAWRA Journal of the American Water Resources Association {\bf{25}}, 349 (1989).
\bibitem{Hendry1999} M.J. Hendry, J.R. Lawrence, and P. Maloszewski, Ground Water {\bf{37}}, 1 (1999). 
\bibitem{Foppen2005} J.W.A. Foppen, A. Mporokoso and J.F. Schijven, J. Contam. Hydrol. {\bf{76}}, 191 (2005). 
\bibitem{Tufenkji2007} N. Tufenkji, Adv. Water. Resour. {\bf{30}}, 1455 (2007).
\bibitem{Bai2016} H. Bai et al. RSC Adv. {\bf{6}}, 14602 (2016).
\bibitem{Lutterodt2009} G. Lutterodt, M. Basnet, J.W.A. Foppen, and S. Uhlenbrook, Water Res. {\bf{43}}, 595 (2009).
\bibitem{Lutterodt2011} G. Lutterodt, J.W.A. Foppen, A. Maksoud, and S. Uhlenbrook, J. Contam. Hydrol. {\bf{119}}, 80 (2011).
\bibitem{Camesano1998} T.A. Camesano and B. E. Logan, Environ. Sci. Technol., {\bf{32}}, 1699 – 1708 (1998)
\bibitem{Becker2003} M.W. Becker et al. Ground Water {\bf{41}}, 682 (2003). 
\bibitem{Tao2009} L. Tao, and R.M. Ford, Environ. Sci. Technol. {\bf{43}}, 1546 (2009).
\bibitem{Rusconi2014} R. Rusconi, J.F. Guasto and R. Stocker, Nat. Phys. {\bf{10}}, 212 (2014).
\bibitem{Altshuler2013} E. Altshuler et al. Soft Matter {\bf{9}}, 1864 (2013).
\bibitem{Hill2007} J. Hill, O. Kalkanci, J. L. McMurry, and H. Koser, Phys. Rev. Lett. {\bf{98}}, 068101 (2007).
\bibitem{Kaya2012} T. Kaya, and H. Koser, Biophys. J .  {\bf{102}}, 1514 (2012);
\bibitem{Marcos2012} Marcos, H.C. Fu, T.R. Powers, and R. Stocker, Proc. Nat. Acad. Sci. USA {\bf{109}}, 4780 (2012)
\bibitem{Mino2018} G. Mino et al. Adv. in Microbiol.,  8, 451-464 (2018).
\bibitem{Morales2015} N. Figueroa-Morales et al. Soft Matter {\bf{11}}, 6284 (2015).
\bibitem{Douarche2009} C. Douarche, A. Buguin, H. Salman, \&  A. Libchaber, Phys. Rev. Lett. {\bf{102}}, 198101 (2009). 
\bibitem{Edelstein2014}A.D Edelstein, et al. Journal of Biological Methods {\bf{1}} (2014).
\bibitem{Fiji} Free-ware image analysis Fiji can be uploaded at : https://imagej.net/Fiji/
\bibitem{Saffman1959} Saffman, P. G. J. Fluid Mech. 6, 321 (1959).
\bibitem{Bear1971} Bear J, Dynamics of Fluids in Porous Media (Amsterdam: Elsevier, 1971).
\bibitem{Uhlenbeck1930} G.E. Uhlenbeck and L. S. Ornstein, Phys. Rev. {\bf{36}}, 823 (1930).
\bibitem{Wu2006} M. Wu et al. Appl. Environ. Microbiol. {\bf{72}}, 4987 (2006).
\bibitem{Potomkin2017} M. Potomkin, A. Kaiser, L. Berlyand, I. Aranson, New Journal of Physics {\bf{19}}, 115005 (2017).
\bibitem{Hornberger1992} G.M. Hornberger, A.L. Mills, and J.S. Herman, Water Resour. Res. \textbf{28}, 915 (1992).
\bibitem{McCaulou1994} D.R. McCaulou, R.C. Bales, and J.F. McCarthy, J. Contam. Hydrol. {\bf{15}}, 1 (1994).
\bibitem{Stumpp2011} C. Stumpp, J.R. Lawrence, M.J. Hendry, and P. Maloszewski. Environ. Sci. Technol. {\bf{45}}, 2116 (2011).
\bibitem{Marchetti2013} M. C. Marchetti et al. Rev. Mod. Phys. {\bf{85}}, 1143 (2013).
\bibitem{Solon2015}A.P. Solon et al. Nature Physics {\bf{11}}, 673 (2015).
\bibitem{Ginelli2010} F. Ginelli, F. Peruani, M. Bar, and H. Chaté. Phys. Rev. Lett., {\bf{104}}, 184502, (2010).
\bibitem{Gregoire2004} G. Gregoire and H. Chat\'e, Phys. Rev. Lett. {\bf{92}}, 025702 (2004); 
\bibitem{Wu2000} X.-L. Wu, A. Libchaber, Phys. Rev. Lett. {\bf{84}}, 3017 (2000); 
\bibitem{Galajda2007} P. Galajda, J. Kleymer, P. Chaikin and R. Austin, J. Bacteriol. {\bf{189}}, 8704 (2007).
\bibitem{Hatwalne2004}Y. Hatwalne, S. Ramaswamy, M. Rao, R.A. Simha, Phys. Rev. Lett. {\bf{92}}, 118101 (2004).
\bibitem{Sokolov2009} A. Sokolov and I.S. Aranson, Phys. Rev. Lett. 103, 148101 (2009). 
\bibitem{Rafai2010} S. Rafa\"i, L. Jibuti, P. Peyla, Phys. Rev. Lett. {\bf{104}}, 098102 (2010).
\bibitem{Saintillan2010} D. Saintillan, Exp. Mech. {\bf{50}}, 1275 (2010).
\bibitem{Lopez2015}H.M. L\'opez, J. Gachelin, C. Douarche, H. Auradou, \& E. Clément, Phys. Rev. Lett. {\bf{115}}, 028301 (2015). 

\end{thebibliography}
\end{document}